\documentclass[iop,apj,tighten,numberedappendix]{emulateapj}
\pdfoutput=1
\usepackage{amsmath,amstext}
\usepackage[breaklinks,colorlinks,citecolor=blue,linkcolor=magenta]{hyperref}
\usepackage[all]{hypcap}
\usepackage[usenames,dvipsnames]{color}
\usepackage{enumitem}

\usepackage{newtxtext}
\usepackage{newtxmath}
\usepackage{bm}

\DeclareMathAccent{\dot}    {\mathalpha}{operators}{'137}
\DeclareMathAccent{\ddot}    {\mathalpha}{operators}{'177}

\newcounter{todocount}
\newcommand{\todofull}[1]{   \noindent   \hyperref[#1]{\refstepcounter{todocount}\textcolor{Aquamarine}{\sf\large[\thetodocount]}}\label{back:#1}
  \expandafter\edef\csname todo@#1\endcsname{\thetodocount}
}

\shorttitle{Diffusion and mixing in globular clusters}
\shortauthors{Meiron \& Kocsis}

\begin{document}
\title{Diffusion and mixing in globular clusters}
\author{Yohai Meiron and Bence Kocsis}
\affil{Institute of Physics, E\"otv\"os University, P\'azm\'any P. s. 1/A, Budapest, 1117, Hungary}

\begin{abstract}
Collisional relaxation describes the stochastic process with which a self-gravitating system near equilibrium evolves in phase space due to the fluctuating gravitational field of the system. The characteristic timescale of this process is called the relaxation time. In this paper, we highlight the difference between two measures of the relaxation time in globular clusters: (i) the diffusion time with which the isolating integrals of motion (i.e. energy $E$ and angular momentum magnitude $L$) of individual stars change stochastically and (ii) the asymptotic timescale required for a family of orbits to mix in the cluster. More specifically, the former corresponds to the instantaneous rate of change of a star's $E$ or $L$, while the latter corresponds to the timescale for the stars to statistically forget their initial conditions. We show that the diffusion timescales of $E$ and $L$ vary systematically around the commonly used half-mass relaxation time in different regions of the cluster by a factor of $\sim 10$ and $\sim$100, respectively, for more than $20\%$ of the stars. We define the mixedness of an orbital family at any given time as the correlation coefficient between its $E$ or $L$ probability distribution functions and those of the whole cluster. Using Monte Carlo simulations, we find that mixedness converges asymptotically exponentially with a decay timescale that is $\sim10$ times the half-mass relaxation time.
\end{abstract}


\maketitle

\section{Introduction}\label{sec:introduction}
Star clusters evolve under many external and internal factors. Externally, depending on the star cluster's location in the host galaxy, gravitational perturbations (in the form of a tidal field), occasional collision with a giant molecular cloud or another star cluster, or a supermassive black hole, all affect the structure of clusters on timescales which are fairly short compared to the age of the universe. Internally (after a short phase in which gas dynamics and violent relaxation dominate, ending in a cluster which is nearly gas free and spherical), stellar evolution causes individual stellar masses to generally decrease and may provide ambient gas for the formation of a second generation of stars. Energy and angular momentum exchange due to gravitational interactions between stars (single, binary, or multiple) leads to a plethora of phenomena such as mass segregation, core collapse, and cluster evaporation \citep[][and references therein]{Spitzer87,Binney+08,Merritt13}. This so-called collisional evolution contrasts with the evolution of systems on galactic or cosmological scales due to the large scale gravitational field. Each of these factors has been studied independently, and all of them in tandem in more modern star cluster modeling, that includes realistic prescriptions for much of the physics involved \citep{Sippel+13,Heggie14,Wang+16}.

Relaxation is the idea at the heart of collisional evolution. It is conceptually useful to split the gravitational potential of the star cluster into an approximately time-independent smooth potential and a fluctuating component which accounts for time-dependent finite number effects (i.e. dynamical two-body encounters and resonant\footnote{Resonant in the sense of the commensurability condition in the mean field potential \citep[e.g.][and reference therein]{Merritt15}} multi-body interactions; see e.g. \citealt{Fouvry+17}). In spherical star clusters, the smooth component allows four independent isolating integrals (energy and three angular momentum vector component) to exist. The fluctuating component drives slow variations in their values, a process in which the 6D phase space distribution function evolves towards the maximum entropy configuration. This effect is described approximately by diffusion (\citealt{Chandrasekhar42} and many references thereafter; cf. \citealt{BarOr+12} who discuss anomalous diffusion in galactic centers).

The modern idea of relaxation was first introduced in thermodynamics by \citet{Maxwell1866}. \citet{Jeans1913} applied this to stellar dynamics by assuming that stars, like molecules in a gas, are subjected to thermal agitation. He estimated the timescale associated with relaxation based on deflection angle and the idea of mean free path. Around the same time, Karl Schwarzschild was working on the velocity distribution in the Galaxy. He investigated how a Maxwellian velocity distribution may be produced in stellar systems. He also derived the timescale for this to occur based on perturbations to the orbital energy due to successive stellar encounters (\citealt{Schwarzschild24}; published posthumously\footnote{The paper fragment (in German) is also found in \citet{Voigt92} with foreword by R. Wielen.}). \citet{Chandrasekhar42} extended those ideas and provided rigorous evaluation of those timescales (based on both deflection angle and energy), his scattering theory will be discussed in more detail in Section \ref{sec:relaxation}. These early authors considered a very simplified model for a star cluster, namely infinite and homogeneous. In this case, the rate of diffusion is of course the same everywhere. More modern kinetic approaches have been developed that account for spatial inhomogeneity and collective effects \citep{Heyvaerts10,Chavanis12,Chavanis13,Sridhar+16}.

Fundamental differences between the inter-molecular forces in gas and the gravitational force (such as its long range and always-attractive nature) lend star clusters very different thermal properties. Nevertheless, models of a star cluster as a gravitating gaseous sphere (analogous to a star; sometimes called fluid-dynamical models) were developed in the late 1970s and were quite successful in investigating core collapse (\citealt{Hachisu+78,Lynden-Bell+80}, and for the original connection see \citealt{Lynden-Bell+68}). More elaborate Fokker--Planck models followed (e.g. \citealt{Cohn79} following from earlier works such as \citealt{Henon61} and \citealt{Kuzmin57}), based on orbit-averaged diffusion coefficients, which required less assumptions than the gaseous models. These more accurate models made it possible to study the time evolution of star clusters, in the continuum (large $N$) limit, with relative modest computational effort. They considered a cluster's inhomogeneous density profile, where now the relaxation time may vary considerably between the inner and outer parts of the cluster.

These previous works have devoted much attention to the collisional evolution of the star cluster as a whole, but not much focus has been given to the evolution of particular orbital families, i.e. stars with similar (initial) values of total energy $E$ and angular momentum magnitude $L$ with respect to the cluster's center. Furthermore, multiple interpretations have been offered for the term \emph{relaxation time}, namely that it is the (mean) time for a quantity to change by order of itself, or that it is the timescale for a star to statistically ``forget its initial conditions''. Under the circumstances relevant for a star cluster, however, those definitions are not the same as the diffusion timescale. While the diffusion time is an instantaneous timescale, those definitions describe a long term behavior we call \emph{mixing} and discuss further and quantify in Sections \ref{sec:diffusion} and \ref{sec:mixing}.

Since non-resonant 2-body relaxation is the main process responsible for both $E$ and $L$ exchange in star clusters, in this paper we will use the term \emph{relaxation time} for the timescales associated with the diffusion of either $E$ or $L$. The direction of the angular momentum vector is also a constant of motion which is affected by 2-body relaxation, in this case, however, vector resonant relaxation \citep{Rauch+96} may play a role as well. The relative importance of vector resonant versus 2-body relaxation in star clusters (specifically, globular cluster lacking a central singularity) will be investigated in a future paper \citep{Meiron+18}.

In this paper, rather than studying the collisional evolution of the star cluster as a whole, we focus on the evolution (in the statistical sense) of particular orbital families, i.e. stars that are initially in a small neighborhood of a point in $(E,L)$-space. We use an idealized isolated Plummer model (which is isotropic) as an example, where all stars have the same mass, their masses are constant in time, and there are no binaries. The two aspects of this study are to find the $E$- and $L$-relaxation times as a function of $E$ and $L$, and to statistically follow representative orbital families in time and quantify their degree of mixing using a quantity we call mixedness.

In Section \ref{sec:diffusion} we discuss diffusion in general terms; in Section~\ref{sec:relaxation} we quantify the relaxation time and calculate it for different orbital families in a Plummer model; in Section \ref{sec:mixing} we discuss the concept of mixing and how it is quantified by mixedness, which we measure for representative orbital families in a Plummer model; finally, we discuss general and astrophysical significance in Section \ref{sec:discussion}.

\section{Short and long term behavior}\label{sec:diffusion}

Diffusion due to 2-body encounters is the dominant cause of change of the energy and angular momentum of individual particles in a system in equilibrium\footnote{Collisionless equilibrium; see chapter 4 of \citet{Binney+08}}. A secondary cause is the gradual change of the global potential (which in our case is indirectly due to 2-body encounters but in the general case could be due to other reasons, e.g. change of the tidal field due to the cluster's motion through a galaxy). A timescale associated with the diffusion time of any constant of motion is commonly referred to as the relaxation time.

The rate of diffusion, expressed by the diffusion coefficients, is not uniform and depends on phase-space coordinates. Since individual particles are generally not stationary and move in phase-space even without diffusion, the concept of a relaxation time is meaningful if it is associated with some kind of averaged diffusion coefficient. Even so, the relaxation time is only an instantaneous timescale \citep{Chandrasekhar42} akin to the local slope of a curve. In this paper, we identify the relaxation time with the instantaneous ensemble-average of the diffusion time of the energy or angular momentum magnitude of individual stars. We calculate the diffusion time as the average initial rate of square change of energy and angular momentum due to two-body encounters with other stars in the cluster.

Mixing of a property $x$ is the tendency of a distribution of $x$ of any subpopulation in the cluster to evolve toward the distribution in the whole cluster (which itself may be changing in time due to collisional evolution), it occurs due to diffusion and therefore is not a separate physical process. This then describes the long term behavior of a system, as opposed to the instantaneous diffusion time. The subpopulation is a set of particles with a very narrow (initial) distribution of $x$, and $x$ in our case is a constant of motion (i.e. a quantity that would not change in the absence of diffusion). In particular, for an approximately spherically symmetric cluster, it is a set of orbits with nearly the same (initial) semi-major axis and eccentricity\footnote{In non-Keplerian spherically symmetric potentials, orbits are planar ``rosettes'' rather than ellipses, but one can still define orbital elements geometrically. The pericenter $r_\mathrm{p}$ and apocenter $r_\mathrm{a}$ are the radial turning points of motion which satisfy $E= \Phi_{\rm eff}(r,L)$, where $\Phi_{\rm eff}(r,L)= \Phi(r) + \frac12 (L/r)^2$ is the effective radial potential, $\Phi(r)=\sum_{i=1}^N Gm/|\bm{r}_i-\bm{r}|$ is the potential, and $E$ and $L$ are respectively the energy and angular momentum per unit mass, which are approximately conserved in an approximately spherical cluster. The semi-major axis and eccentricity are defined as $a=\frac{1}{2}(r_\mathrm{p}+r_\mathrm{a})$ and  $e=(r_\mathrm{a}-r_\mathrm{p})/(2a)$.} (but arbitrary orientations of the orbital plane, orbital phase etc.) Consider the energy (e.g.) distribution of an orbital family: it is initially very narrow but widens with time due to diffusion. We expect that it will approach asymptotically to the energy distribution of all particles in the system. We quantify the amount of mixing using a quantity called mixedness, defined in Section~\ref{sec:mixing}.

In the following sections we explore the concepts of relaxation and mixing in a more detailed way, using the Plummer model to illustrate each one.

\section{Relaxation time}\label{sec:relaxation}

\subsection{Basic concepts}\label{sec:relaxation:basic}

The following formula is often used to estimate the relaxation time in a stellar system
\begin{equation}
t_{\mathrm{relax}}=\frac{\alpha\sigma^{3}}{G^{2}m\rho\ln\Lambda}\label{eq:chandra}
\end{equation}
where $\sigma\equiv\sqrt{\langle v^2 \rangle/3}$ is the one-dimensional velocity dispersion, $m$ is a particle's mass, $\rho$ is the particle mass density, and $G$ is the gravitational constant. The dimensionless factors $\alpha$ and $\ln\Lambda$ hide much of the complexity of the problem, they vary depending on the exact definition and will be discussed below. This formula has proved quite useful but it is important to understand its caveats. Derivation of this type of formula \citep{Chandrasekhar42,Spitzer87,Binney+08} requires making several assumptions. The first and perhaps most critical one is that diffusion can be adequately described by a superposition of independent 2-body interactions. Additional assumptions include the uniform spatial density (which implied the neglect of self-gravity), the isotropy of velocity field, and the that the velocities follow the Maxwell--Boltzmann distribution characterized by $\sigma$. Those assumptions are really valid only in a hypothetical infinite and homogeneous medium, where the mean-field gravity is neglected. To apply this to the case of star clusters as an average, global quantity, additional approximations are made, namely that the density is equal to the average density within the half-mass radius $r_\mathrm{h}$, and that $\sigma=\beta\sqrt{GmN/r_\mathrm{h}}$ (based on the virial theorem, with $\beta$ an order unity constant). The result is what is commonly called the \emph{half-mass relaxation time}
\begin{equation}
t_{\mathrm{rh}} = \frac{\gamma}{\ln\Lambda} \sqrt{\frac{N r_\mathrm{h}^3}{Gm}} = \frac{\gamma N t_\mathrm{dyn}}{\ln\Lambda}\label{eq:trh}
\end{equation}
where $\gamma=8\pi\alpha\beta^3/3$, $N$ is the total number of stars in the cluster, and $t_\mathrm{dyn}^2 \equiv r_\mathrm{h}^3/GM$ is the dynamical time ($M=mN$ is the total cluster mass). \citet{Spitzer87} got a value of $\gamma=0.138$ from simple considerations. In the following sections we will use $t_{\mathrm{rh}}$ as a reference time since it is a very simple estimate to make, and despite the many approximations, it retains the correct scaling with the number of particles. Thus, the results discussed throughout this paper are independent of $N$.

\citet[and references therein]{Chandrasekhar42} proposed a definition for the relaxation time based on energy, leading to an expression of the form of Equation~(\ref{eq:chandra}), where the relaxation time is said to have been reached when the cumulative square change of energy $(\Delta E)^2$ becomes of the same order as the square of the initial kinetic energy. The average $\langle(\Delta E)^2\rangle$ per unit time (the diffusion coefficient) is meticulously calculated by considering the root-mean-square (rms) energy change due to 2-body encounters sampled independently from a homogeneous uniform medium within a given minimum and maximum impact parameter and a Maxwell--Boltzmann velocity distribution. This gives $\alpha = 9/(16\sqrt{\pi}) \approx 0.317$.

The approach taken by \citet{Spitzer87}\footnote{Both Chandrasekhar and Spitzer give one additional definition each for the relaxation time. The former's is based on the deflection angle rather than energy change, and the latter's is the time-dimensioned constant of the encounter term in the Fokker--Planck equation.} and \citet[and references therein]{Binney+08} was essentially identical with respect to summing up individual encounters, but they computed the change in velocity components during the encounters, rather than energy. Since as noted above, the medium is approximated as infinite and homogeneous, energy and velocity magnitude are interchangeable\footnote{If the mean field is spatially homogeneous, the three velocity components are three integrals of motion, which change due to the stochastic fluctuating component of the potential.}. The mean change in the parallel component of the velocity $\Delta v_\parallel$, its square $(\Delta v_\parallel)^2$ and the square change of the perpendicular component $(\Delta v_\bot)^2$ are calculated per unit time to yield diffusion coefficients. One advantage in this method is that it gives the expression for dynamical friction for free through the so-called drift term $\langle\Delta v_\parallel\rangle$. Another advantage is that up to second order, the diffusion of coefficients of any quantity (such as energy) can be written as a linear combination of the velocities. We make use of this property to derive the local angular momentum diffusion coefficients in Appendix~\ref{sec:local-diff-L}. An additional difference to Chandrasekhar's approach is that instead of assuming a foreground velocity distribution equal to the background velocity distribution and averaging over both, these authors more simply substitute a typical value for the velocity for the test star, which gives a value of $\alpha\approx 0.340$ (the exact value can be written as a complicated expression involving the error function).

The factor $\ln \Lambda\approx \ln(b_\mathrm{max}/b_\mathrm{min})$ is the Coulomb logarithm which crops up in the derivation due to the divergence of the integral over impact parameters. This divergence occurs on small scales due to the fact that the small-angle deflection approximation mishandles strong collisions, and on large scale due to the local approximation (i.e. the neglect of inhomogeneity). Chandrasekhar interpreted $b_\mathrm{max}$ as the inter-particle distance, but it was pointed out by \citet{Cohen+50} that it should be the order of the size of the system (or the region that contains most of the particles). Due to the uncertainty, in star clusters investigations $\ln\Lambda$ is generally set to $\ln(\lambda N)$ where the value of $\lambda$ could be empirically determined from $N$-body simulations (\citealt{Giersz+94} got $\lambda\approx 0.11$).

\citet{Larson70a} derives a timescale for the collisional evolution in a substantially different way. His work is based on reorganizing the Fokker--Planck equation as a set of moment equations. The velocity moments, representing different kinds of deviations from a Maxwellian velocity distribution, are shown to decay exponentially with timescales similar (i.e. up to an order unity factor) to the relaxation time defined by Chandrasekhar (Equation \ref{eq:chandra}), with an apparent tendency for the higher moments to relax more slowly than the lower ones.

While Equation~(\ref{eq:trh}) is a very useful timescale parameter for a star cluster, it hides the very important information of how the diffusion timescale depends on the location within the cluster (e.g. the central regions versus the outskirts). Equation~(\ref{eq:chandra}) is a bit more general and one could in principle substitute for a spherically symmetric star cluster $\sigma(r)$ and $\rho(r)$ corresponding to a particular model and obtain an expression for the relaxation time which is a function of radius. This is somewhat an abuse of Equation~(\ref{eq:chandra}) as it is derived under the assumption that $\sigma$ and $\rho$ are constants and the medium is infinite. Also, this kind of calculation will not yield the dependence on the test star's eccentricity. In the two following subsections, we calculate the relaxation time (for both energy and angular momentum) for each orbital family in a Plummer model by using Chandrasekhar's basic scattering theory, but not making further assumptions apart from the isotropy of the model (which is justified for a Plummer model).

\subsection{Methods}\label{sec:relaxation:methods}
We calculate the $E$ and $L$-diffusion coefficients and the corresponding relaxation timescales for a specific $(E,L)$ orbital family due to stellar scattering using the local diffusion approximation. This amounts to adding up the contributions of incoherent local two-body flyby encounters, assuming that (i) the flyby events have a short duration relative to the orbital timescale and (ii) that the encounters are predominantly local, where the density of scatterers is approximately homogeneous. The changes in $E$ and $L$ are accumulated incoherently over the unperturbed orbit in the cluster with a given $E$ and $L$, and the relaxation times follow from there.

\citet{Rosenbluth+57} gave implicit expressions for the average change per unit time of a test particle's velocity components $\langle \Delta v_i \rangle$, as well as $\langle \Delta v_i \Delta v_j \rangle$, due to 2-body encounters, which depend on the target particle's velocity and the background particles' velocity distribution. Under the assumption of an isotropic velocity field, these expressions could be simplified to yield three useful functions expressing the average change parallel to the original direction of velocity $\langle \Delta v_\parallel \rangle$ as well as the square change in the parallel direction $\langle (\Delta v_\parallel)^2 \rangle$ and perpendicular $\langle (\Delta v_\bot)^2 \rangle$ to the original velocity vector (note that isotropy of the velocity field necessitates $\langle \Delta v_\bot \rangle=0$). Let us remember that these are changes per unit time, despite the notation.

Given these functions, we can express (up to second order) the average square change per unit time of the energy and angular momentum
\begin{align}
\left\langle (\Delta E)^{2}\right\rangle  &=v^{2}\left\langle (\Delta v_{\parallel})^{2}\right\rangle \label{eq:local-diff-E} \\
\left\langle (\Delta L)^{2}\right\rangle  &=\frac{r^{2}}{v^{2}}\left[v_{t}^{2}\left\langle (\Delta v_{\parallel})^{2}\right\rangle +\textstyle{\frac{1}{2}}v_{r}^{2}\left\langle (\Delta v_{\bot})^{2}\right\rangle \right] \label{eq:local-diff-L}
\end{align}
where $r$ is the radial coordinate (the cluster center is at the origin of the coordinate system), $v_r$ is the velocity component in the radial direction, $v_t$ is the tangential velocity component, and $v^2=v_r^2+v_t^2$. Equation~(\ref{eq:local-diff-E}) is very simple to derive: one just needs to remember that due to assumption (i) above (short duration encounters), the potential energy does not change during an encounter, and velocity terms with powers higher than two are neglected \citep[equation 2-51 therein]{Spitzer87}. Equation~(\ref{eq:local-diff-L}), on the other hand, is more difficult to derive. The full derivation is given in Appendix~\ref{sec:local-diff-L}.

In Appendix~\ref{sec:orbital-averaging} we substitute Rosenbluth's expressions for $\langle (\Delta v_\parallel)^2 \rangle$ and $\langle (\Delta v_\bot)^2 \rangle$ (both scale linearly with density and are otherwise functions of $v$ only) and perform orbital averaging. The result is two functions denoted by $D_{E^2}(E,L)$ and $D_{L^2}(E,L)$ called the orbitally averaged diffusion coefficients. \citet{Cohn+78} have similarly derived the diffusion coefficients, but instead of $L$ they considered the square relative angular momentum and performed orbital averaging under the assumption of Keplerian orbits \citep[see also][]{Merritt15}.

It is natural to define the relaxation time for any property $x$ simply as $x^2/D_{x^2}$, but this may lead to strange results. For example, if we defined the $E$-relaxation time as $E^2/D_{E^{2}}$, it would \emph{decrease} from the center of the cluster outward, which goes against the intuition that diffusion is more important in the innermost regions. Further on physical grounds, we may note that the gravitational scattering process, assuming local short-duration encounters, changes the instantaneous velocity directly, hence the instantaneous kinetic energy, while the potential energy with respect to the cluster is fixed. Therefore we define the energy-relaxation time specifically with respect to the average \emph{kinetic} energy for a star on an $(E,L)$ orbit $\overline{E_\mathrm{k}}$. While the kinetic energy is not a constant of motion, its orbital average (by definition) is. Unlike in the Keplerian case where $\overline{E_\mathrm{k}}=|E|$, in star clusters $\overline{E_\mathrm{k}}$ strongly differs from this result especially in the inner regions of the cluster where it is much smaller than $|E|$. It generally depends on both $E$ and $L$ through the potential.

A related issue for $L$-relaxation, if it were defined as $L^2/D_{L^2}$ then any nearly radial orbit would have asymptotically zero relaxation time. Indeed, if the direction of angular momentum vector is nearly a null-vector, its direction physically changes by an arbitrarily high rate for nearly radial orbits due to any finite torque. However, since we are mainly interested in the long term evolution of the magnitude of angular momentum vector, we choose the reference angular momentum to be the maximal (circular) angular momentum $L_{\mathrm{c}}$ corresponding to the given energy of the orbit. Using this definition, the $L$-relaxation time is simply proportional to the inverse net torque exerted on the orbit due to two-body encounters, which is finite even for $L=0$. This definition represents an upper limit of the actual angular momentum diffusion time.

Thus, the relaxation times are defined as follows
\begin{align}
t_{\mathrm{rx},E} &\equiv \overline{E_\mathrm{k}}^2/D_{E^{2}}\\
t_{\mathrm{rx},L} &\equiv  L_{\mathrm{c}}^{2}/D_{L^{2}}
\end{align}
Where except $L_{\mathrm{c}}$ which is only a function of $E$, all other variables are functions of both $E$ and $L$.

\subsection{Results}\label{sec:relaxation:results}

\begin{figure}
\includegraphics[width=1\columnwidth]{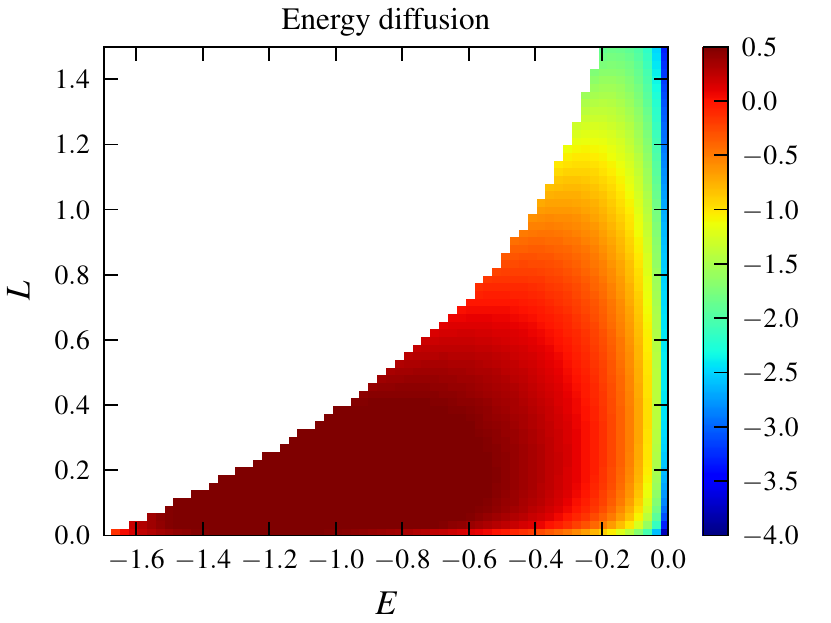}\label{fig:trxE}
\caption{Energy diffusion coefficient for a Plummer model, calculated by orbital averaging for each point on a grid in $(E,L)$ space. The color scale is $\log_{10}[ N D_{E^2}/(\ln\Lambda \langle E^2\rangle)]$ where $N$ is the number of particles, $\ln\Lambda$ is the Coulomb logarithm, and $\langle E^2\rangle)$ is the mean square energy of particles in a Plummer model. This normalization guarantees dimensionlessness and independence of the number of particles. The axes are in H\'enon units for a Plummer model with virial radius of one H\'enon length unit (see text for details).}\label{fig:relaxation-E}
\end{figure}

\begin{figure}
\includegraphics[width=1\columnwidth]{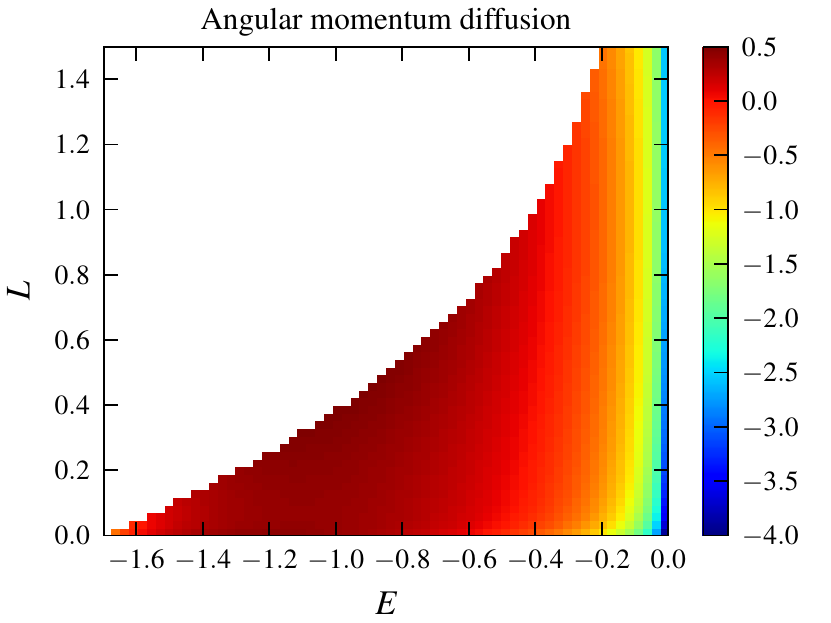}
\caption{Same as Figure \ref{fig:trxE}, but for the angular momentum diffusion coefficient.}\label{fig:relaxation-L}
\end{figure}

\begin{figure}
\includegraphics[width=1\columnwidth]{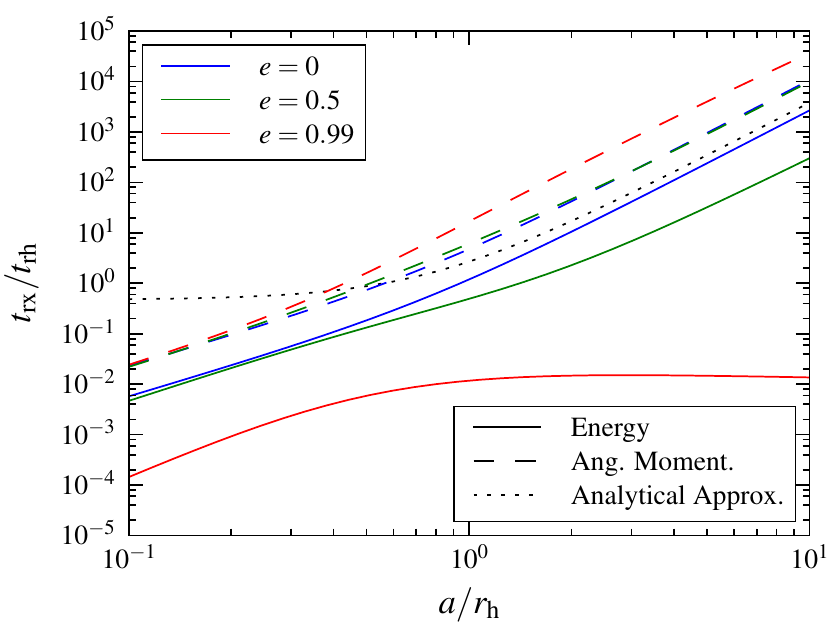}
\caption{The energy (solid lines) and angular momentum (dashed lines) relaxation times as function of the semi-major axis for different eccentricity cases in a Plummer model. The blue, green, and red lines represent zero, moderate ($e=0.5$) and high ($e=0.99$) eccentricity, respectively. The dotted black line represents the analytical approximation in Equation (\ref{eq:chandra}). The times are normalized by $t_\mathrm{rh}$ (given by Equation \ref{eq:trh}) while the semi-major is normalized by the half-mass radius.}\label{fig:relaxation-elements}
\end{figure}

\begin{figure}
\includegraphics[width=1\columnwidth]{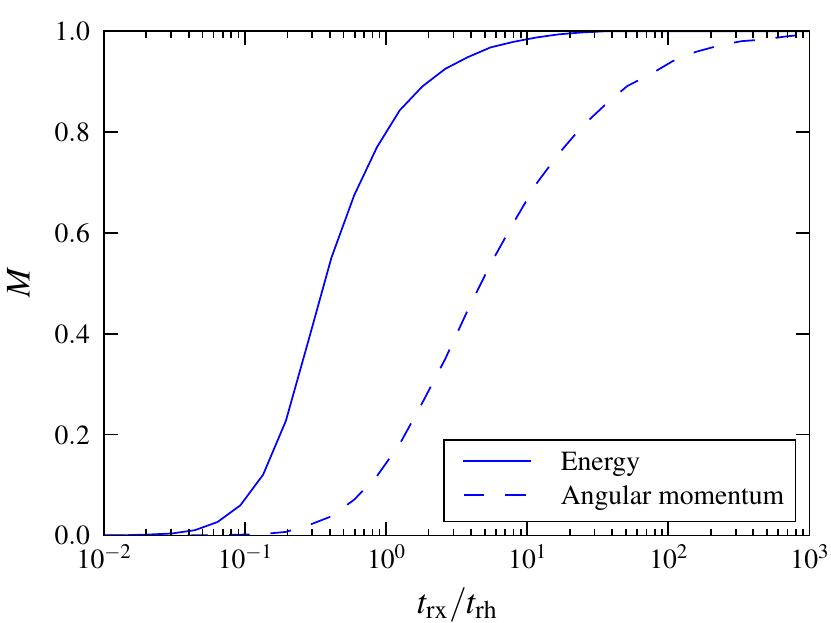}
\caption{The cumulative mass normalized to the total cluster mass (or cumulative number of stars normalized to the total number of stars) with energy (solid line) and angular momentum (dashed line) relaxation times shorter than $t_\mathrm{rx}$ in units of $t_\mathrm{rh}$. }\label{fig:relaxation-cumulative}
\end{figure}

In Figures \ref{fig:relaxation-E} and \ref{fig:relaxation-L} we present the variation of the diffusion coefficients in  $(E,L)$-space for a Plummer model (see Appendix \ref{sec:orbital-averaging}). The units of $E$ and $L$ in the figures are given in H\'enon units\footnote{Also known as $N$-body units, where the model's total mass and the gravitational constant $G$ are set to unity, and the total energy of the model is set to $-1/4$. For a cluster with total mass $M$ and Plummer radius $r_0$ (specified in whatever physical units), the H\'enon energy and angular momentum (per unit mass) units are $[E]=\frac{3\pi}{16} \frac{GM}{r_0}$ and $[L]=\sqrt{\frac{16}{3\pi} GMr_0}$, respectively.} for a Plummer model that is normalized such that its virial radius is unity (giving a Plummer radius of $3\pi/16$). The half-mass radius in this case is $r_\mathrm{h} \approx 0.7686$ H\'enon length units. The diffusion coefficients are presented on a logarithmic scale as dimensionless quantities which are independent of the number of particles or choice of Coulomb logarithm. In order to achieve that, we normalize the diffusion coefficients by multiplying them by
\begin{equation}
\frac{N}{\langle x^2 \rangle \ln\Lambda}
\end{equation}
where $x$ is either $E$ or $L$ and $\langle x^2 \rangle$ indicates its mean square for the entire cluster. We can obtain those cluster averages by performing the appropriate integrals on the distribution function (Equation \ref{eq:plummer-df}). In H\'enon units for our given model, $\langle E^2 \rangle = 704/(105\pi^2)\approx0.68$ and $\langle L^2 \rangle = 9\pi^2/256 \approx 0.35$.

We can get an analytical approximation of the relaxation time as a function of radius by substituting $\sigma(r)$ and $\rho(r)$ for a Plummer model in Equation~(\ref{eq:chandra}). These are given by \citep[e.g.][]{Binney+08}
\begin{align}\label{eq:sigma}
\sigma^{2}(r) =\frac{GM}{6r_{0}}\left[1+\left(\frac{r}{r_{0}}\right)^{2}\right]^{-1/2}\\\label{eq:rho}
\rho(r)       =\frac{3M}{4\pi r_{0}^{3}}\left[1+\left(\frac{r}{r_{0}}\right)^{2}\right]^{-5/2}
\end{align}
It appears as the dotted black line in Figure \ref{fig:relaxation-elements}. This figure also shows the relaxation times calculated numerically according to Section \ref{sec:relaxation:methods} as a function of semi-major axis for various eccentricity values. The energy and angular momentum relaxation times are represented by the solid and dashed lines, respectively, while the color represents the eccentricity. Circular orbits (blue), intermediate eccentricity of $e=0.5$ (green) and high eccentricity of $0.99$ (red). The relaxation time is normalized by $t_\mathrm{rh}$ given by Equation (\ref{eq:trh}) so it is independent of $N$ and $\ln\Lambda$, while the semi-major axis is normalized by the half-mass radius $r_\mathrm{h}$.

In particular, for a circular orbit with $a=r_\mathrm{h}$ the energy relaxation time $t_{\mathrm{rx},E}=1.2t_\mathrm{rh}$ while the angular momentum relaxation time $t_{\mathrm{rx},L}=4.8t_\mathrm{rh}$ and the analytical approximation gives $2.7t_\mathrm{rh}$. As noted above, the dependence on $N$ and $\ln\Lambda$ is normalized out of these results by presenting them in units of $t_\mathrm{rh}$, but the exact numbers do depend on the particular choice of $\gamma$, and for the analytical approximation, on $\alpha$ as well. One may in fact go backward from those results and tune the dimensionless parameters of $t_\mathrm{rh}$ to get a better analytical estimate for a Plummer sphere. Generally, Figure \ref{fig:relaxation-elements} shows that the approximation using Equations~(\ref{eq:sigma})--(\ref{eq:rho}) overestimates the relaxation time for small $a$. For large $a$, the dependence of both $t_{\mathrm{rx},L}$ and $t_{\mathrm{rx},E}$ on $a$ is the same as the analytical approximation's dependence on radius (with the exception of $t_{\mathrm{rx},E}$ in the $e=0.99$ case) namely $\sim r^{7/2}$. For the highest eccentricity case shown, the $E$-relaxation time seem to be independent of $a$ for $a \gtrsim r_\mathrm{h}$. For large values of $a$, the analytical approximation underestimates $t_{\mathrm{rx},L}$ but overestimates $t_{\mathrm{rx},E}$. This discrepancy is only $\sim 30\%$ for circular orbits but more than a factor of 10 in the moderately eccentric case.

Figure \ref{fig:relaxation-cumulative} shows the cumulative mass with energy (solid line) and angular momentum (dashed line) relaxation times \emph{shorter than} $t_\mathrm{rx}$. Half of the stars in the cluster have energy relaxation time shorter than $0.37t_\mathrm{rh}$, while 90\% have energy relaxation time shorter than $2.0t_\mathrm{rh}$. The corresponding numbers for the angular momentum relaxation time are $4.7t_\mathrm{rh}$ and $58t_\mathrm{rh}$, respectively. This shows that the relaxation times are broadly distributed. Angular momentum magnitude diffusion is systematically slower by an order of magnitude than energy diffusion, which is mostly due to our particular definition of diffusion times e.g. normalized to the average kinetic energy and circular angular momentum, respectively.

\section{Mixing}\label{sec:mixing}

\subsection{Representative regions}\label{sec:mixing:regions}

To investigate mixing in a star cluster, we choose four representative initial orbital families and follow their collisional evolution. These four families specified in Table \ref{tab:regions}, correspond to very small regions in $(E,L)$-space. In each region, the energy and angular momenta fall between $E\pm \Delta E$ and $L\pm\Delta L$, respectively, and they are centered at semi-major axis and eccentricity given in Table \ref{tab:regions}.
\begin{enumerate}[label=(\Roman*)]   \item \label{I} represents the inner region, low energy and intermediate (relative) angular momentum particles. The energy range is selected so that approximately 90\% of particles have higher energy than the middle of the range, the angular momentum range is selected so that the corresponding eccentricity is around 0.5.   \item \label{II} represents the intermediate region with the most typical particles in the system, in the sense that the middle of the range is selected close to the geometric median of all $(E,L)$ values.   \item \label{III} represents traversing orbits between the outer and inner regions with high energy and low relative angular momentum. The energy range is selected so that approximately 90\% of particles have lower energy than the middle of the range, the angular momentum range is selected so that the corresponding eccentricity is around 0.9.   \item \label{IV} represents the outer region, high energy and high relative angular momentum. It has the same energy range as region \ref{III} but the the angular momentum range is selected so that the corresponding eccentricity is lower than 0.3 (in all other regions the semi-major axis as well as the eccentricity have narrow distributions).
\end{enumerate}
In the two following subsections we describe the Monte Carlo simulations we performed, and how we used them to follow the widening of the $E$ and $L$ probability distributions of particles in these regions in time to investigate the long term collisional behavior.

\begin{table}
\caption{Four representative initial regions of orbits in $(E,L)$ space examined for mixing in a Plummer model in Section \ref{sec:mixing}. The corresponding semi-major axis $a$ and eccentricity $e$ for the $(E,L)$ values are given to one digit accuracy, while in the the case of region \ref{IV} the spread of eccentricities is larger than in the other regions. $E$, $L$, and $a$ are in H\'enon units for a Plummer model with virial radius of one (see text for details). The relaxation times for the energy and angular momentum are calculated according to the procedure described in Section \ref{sec:relaxation:methods} and are given in units of $t_\mathrm{rh}$ (Equation \ref{eq:trh}).}\label{tab:regions}
\begin{center}
\begin{tabular}{|c|c|c|c|c|c|c|c|c|c|c|c|}
\hline Region & $E$     & $L$  & $\Delta E$ & $\Delta L$ & $a$ & $e$ & $t_{\mathrm{rx},E}$ & $t_{\mathrm{rx},L}$\\
\hline \hline (I)    & $-1.21$ & 0.16 & 0.01 & 0.01 & 0.2 & 0.5 & 0.1 & 0.8\\
\hline (II)   & $-0.78$ & 0.41 & 0.01 & 0.01 & 0.4 & 0.5 & 0.4 & 4.0\\
\hline (III)  & $-0.30$ & 0.30 & 0.02 & 0.02 & 1.0 & 0.9 & 0.3 & 83\\
\hline (IV)   & $-0.30$ & 1.20 & 0.02 & 0.02 & 1.0 & $\lesssim0.3$ & 14 & 64\\
\hline
\end{tabular}
\end{center}
\end{table}

\subsection{Methods}

In order to follow statistically the collisional evolution of the four selected orbital families, we performed a series of simulations using the \textsc{mocca} code (MOnte Carlo Cluster simulAtor; \citealt{Giersz+13}). This code is based on the orbit-averaged Monte Carlo method of \citet{Henon71} that was later substantially improved by \citet{Stodolkiewicz86}. The basic idea is that changes in each star's energy and angular momentum from one state to the next (successive states of the system are separated by a time step which is a fraction of $t_\mathrm{rh}$) are computed by randomly selecting the position of the star on its orbit, randomly choosing another star, letting the two interact, and multiplying the effect by an appropriate factor. While \textsc{mocca} is a very sophisticated code, capable of realistically simulating globular clusters including physical effects such as stellar evolution and accurate integration of few-body subsystems, we turned most of these features off and integrated very basic models using the code's dynamics capabilities only. Our models were 1024 separate particle realizations (with different random seeds) of a Plummer model with 16k particles each ($\mathrm{k}=1024$). The models were evolved for about $10t_\mathrm{rh}$.

The relatively large number of models is required since as noted in the previous subsection, we look at very small regions of $(E,L)$-space. The fraction of particles in these regions is as low as $5\times10^{-4}$ (for region \ref{III}). Thus for an $N=16\mathrm{k}$ model, only a handful of particles per model have the desired initial $E$ and $L$ values. We therefore superimpose the particle population at each region from 1024 such simulations. Also, to increase the statistics of the background, we superimpose 64 of these simulations for a total of 1M particles ($\mathrm{M}=2^{20}$), at each snapshot. Despite the large number of models, this type of simulation is computationally inexpensive by modern standards, and the whole model set can be run on a desktop computer within less than a day.

\subsection{Results}
\begin{figure*}
\includegraphics[width=504pt]{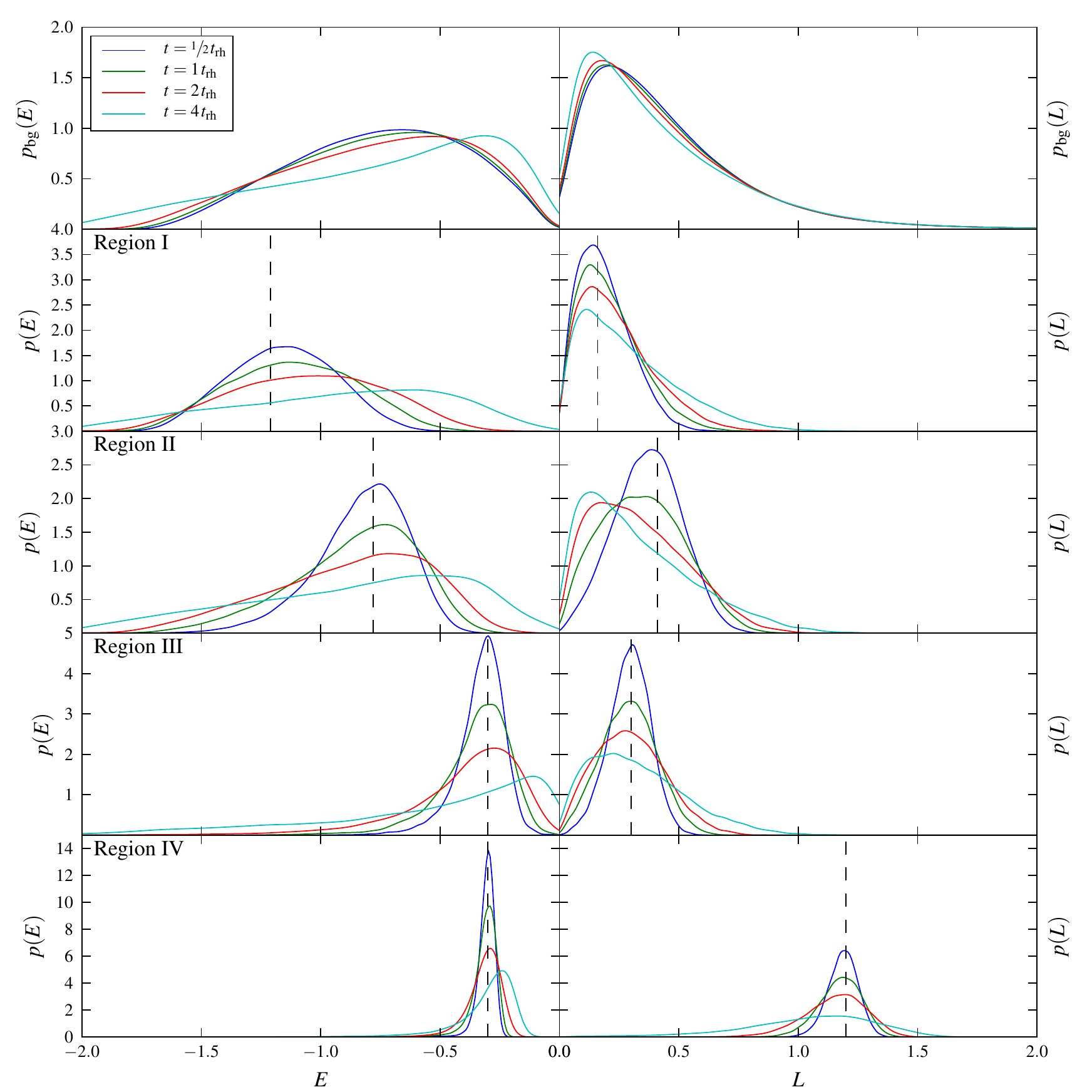}
\caption{The top panels show the evolution of the energy probability distribution (PDF; left) and angular momentum PDF (right) of the whole system due to collisional relaxation and the gravitational response. The four rows of panels below show the evolution of the four representative regions (see Table \ref{tab:regions}). For clarity, only four time epochs are shown: the dashed black line represents the initial energy or angular momentum value (the distributions at $t=0$ resemble Dirac delta-functions), while the blue, green, red and cyan are respectively the distributions at 0.5, 1, 2, 4 times the half-mass relaxation time. }\label{fig:evolution}
\end{figure*}

The top panels of Figure \ref{fig:evolution} show the evolution of the energy probability distribution\footnote{Note that \emph{probability distribution function} is different from the \emph{distribution function} (\textsc{df}). The latter is defined in such a way that $f(\bm{u})\mathrm{d}^{3}\bm{x}\mathrm{d}^{3}\bm{v}$, where $\bm{u}$ is a combination (or several combinations) of the phase space coordinates, is the number of particles (or the mass) inside the 6-dimensional cube of volume $\mathrm{d}^{3}\bm{x}\mathrm{d}^{3}\bm{v}$ around any point $(\bm{x},\bm{v})$ in phase space corresponding to $\bm{u}$. In contrast, the $p(\bm{u})\mathrm{d}\bm{u}$ (where $p(\bm{u})$ is the PDF of some the quantity) is the fraction of particles in the range $\mathrm{d}\bm{u}$ around $\bm{u}$ for any $\bm{x}$ and $\bm{v}$.} (PDF; left) and angular momentum PDF (right) of the whole system due to collisional relaxation and the gravitational response. The four rows of panels below show the evolution of the four representative regions. In the language of quantum mechanics, those graphs show different projections of the system's propagator, which is the probability amplitude for a particle to transition from one state to another in a given time. For clarity, only four time epochs are shown: the dashed black line represents the initial energy or angular momentum value (the distributions at $t=0$ resemble Dirac delta-functions), while the blue, green, red and cyan are respectively the distributions at 0.5, 1, 2, 4 times the half-mass relaxation time defined in Equation (\ref{eq:trh}).

In all cases we find that the $E$- and $L$-distribution of each of the four regions asymptotically approach the system's distribution, which is itself slowly changing in time, and thus they mix toward a fully mixed state. To quantify the degree of mixing, we define \emph{mixedness}, denoted $c$, of a subpopulation with respect to $\bm{u}$ through the correlation coefficient of its PDF $p(\bm{u})$ with that of the fully mixed configuration $p_{{\rm bg}}(\bm{u})$, as
\begin{equation}
c=\frac{\langle p,p_{{\rm bg}}\rangle}{\left\Vert p\right\Vert \left\Vert p_{{\rm bg}}\right\Vert }.\label{eq:mixednessdef}
\end{equation}
Here $\langle A,B\rangle\equiv\int A(\bm{u})B(\bm{u})\mathrm{d}\bm{u}$ is the scalar product on the space of PDFs and $\left\Vert A\right\Vert \equiv\sqrt{\langle A,A\rangle}$. In this paper we restrict our attention to mixing in one dimension only. The PDFs are one dimensional and are denoted by $p(E)$ and $p(L)$, for energy and angular momentum, respectively. The correlation coefficient between distributions varies between zero and unity. $c=0$ represents the completely uncorrelated case, where $p(\bm{u})$ and $p_{{\rm bg}}(\bm{u})$ are have disjoint support sets (in practice this is the case only when $p(\bm{u})$ is a Dirac delta function). $c=1$ represents the fully correlated case where $p(\bm{u})$ and $p_{{\rm bg}}(\bm{u})$ are proportional.\footnote{This number also characterizes the distance between the normalized distributions in the sense that $c=1-\frac{1}{2}\left\Vert \tilde{p}-\tilde{p}_{\mathrm{bg}}\right\Vert ^{2}$ where the tilde denotes $\tilde{A}\equiv A/\left\Vert A\right\Vert $.}

The motivation for this definition is based on a stochastic random-walk model of relaxation introduced by \citet{Kocsis+15}. In that model, each star's actions change randomly in each time step according to a given transition probability function. In that case, it can be shown that the evolution is governed by a linear operator. Decomposing the PDFs of the actions in the orthonormal eigenfunctions of this linear operator shows that each such mode decays independently exponentially in time with distinct decay constants. There is one mode whose decay constant is zero, which represents the fully mixed steady state distribution. The projection of the PDF on the steady state distribution given by Equation (\ref{eq:mixednessdef}) is the natural way to define mixedness in such models.

Estimating $c$ from discrete data can be difficult. One has to first estimate the continuous functions $p$ and $p_{\mathrm{bg}}$ from two discrete sets of values. This could be done by a variety of methods such as kernel density estimation and clustering analysis, most of these have one free parameter or more. Here again we turn to the simplest approach, in this case data binning (histogram). The freedom in this method is to choose the size and position of all the bins. When the number of data points in both sets is extremely large, it is expected that one can produce smooth and fiducial PDFs (with any reasonable density estimation method), but from numerical experiments we found that the relevant dataset sizes are not large enough. A second problem is that even a small bias in the estimation of $c$ may lead to a big systematic error in derivation of a timescale when analyzing the dependence of $c$ on time due to the asymptotic approach to unity as the stellar system is evolving toward a fully mixed state. In Appendix \ref{sec:mixedness-estimation} we describe the numerical procedure to estimate $c$.

\begin{figure}
\includegraphics[width=1\columnwidth]{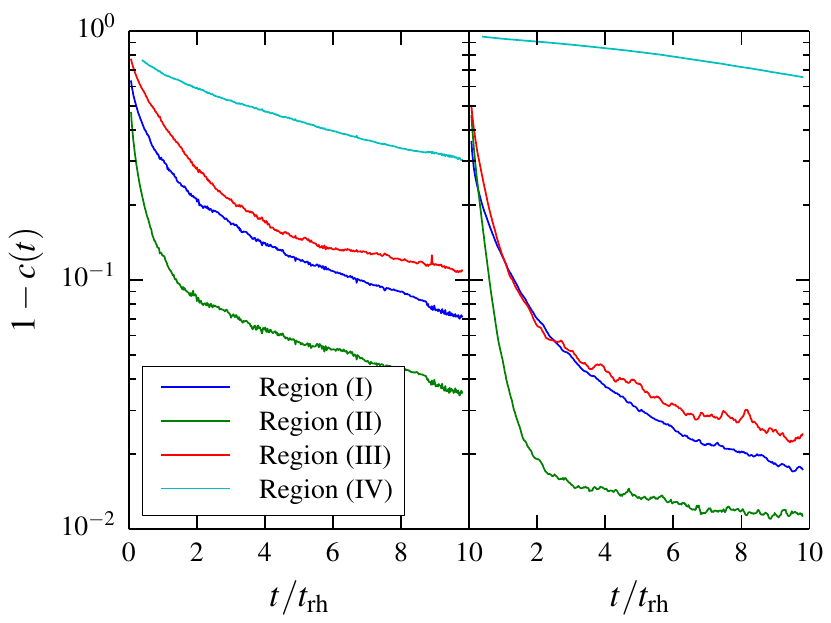}
\caption{One minus the energy (left panel) and angular momentum mixedness (right panel) as a function of time of the four representative orbital families in the $(E,L)$-space described in Section \ref{sec:mixing:regions}.}\label{fig:mixedness}
\end{figure}

Figure \ref{fig:mixedness} shows $1-c(E)$ (left panel) and $1-c(L)$ (right panel) for the four regions. It is evident that the level of mixedness approaches unity asymptotically exponentially in time. Due to the fact that in any real star cluster the number of particles is finite, and the number of particles in any small region of $(E,L)$-space is likewise small, it is expected that the target distribution becomes statistically indistinguishable from the background in a finite amount of time. While the relaxation time is shorter for region \ref{I}, it is evident that region \ref{II} mixes earlier than the others both in energy and in angular momentum. Another conclusion from this figure is that the value of $c(L)$ initially approaches unity faster than $c(E)$ for regions \ref{I}, \ref{II}, and \ref{III}. The opposite happens for region \ref{IV}. This may also be in part related to the proximity of the initial $L$ values to the system's median. Note that the angular momentum mixedness curve for region \ref{II} may be saturating due to the numerical problem with the estimator described above; the relative error in $1-c$ may be very large when $c$ approaches unity.

It is not easy to measure a timescale from the noisy mixedness curves of Figure \ref{fig:mixedness}. One possibility to do so is to choose a threshold (e.g. 90\% or 99\%) and define the mixing time as the time at which the curve (or an extrapolation of which) crosses that threshold. Another, way is to assume that the mixedness curves approach unity exponentially, and define the mixing time as the decay time of this exponential function. Indeed, from the left panel we see that for all curves, $\log_{10}[1-c(t)]$ is roughly linear after about $t=4t_{\mathrm{rh}}$, moreover and all four curves are roughly parallel, implying a shared underlying mixing timescale. By fitting $C$ and $t_{\rm mix}$ assuming $c=1-C\exp(-t/t_{\rm mix})$ we find that the energy mixing time is between $t_{\mathrm{mix},E}\sim9t_{\mathrm{rh}}$ (for region \ref{I}) and $15t_{\mathrm{rh}}$ (for region \ref{III}). The exact values depend on the time interval where the fit is made, and are constrained to within $\sim20\%$. A similar picture is seen in the right panel regarding the angular momentum. In this case the green curve seems to almost stall (implying very long e-folding time). However this is possibly attributed to the numerical problem mentioned above. The timescale derived for regions \ref{I} and \ref{III} is $t_{\mathrm{mix},L}\sim9t_{\mathrm{rh}}$ and $\sim20t_{\mathrm{rh}}$ for region \ref{IV}.

\section{Discussion}\label{sec:discussion}

\subsection{Relaxation}

In Section \ref{sec:relaxation:results} we used scattering theory to calculate the energy $E$ and angular momentum $L$ relaxation times for different orbital families in a Plummer sphere. We showed that the half-mass relaxation time $t_{\mathrm{rh}}$ (Equation \ref{eq:trh}) gives a decent estimate for the order of magnitude of the relaxation times for orbits with semi-major axis that equals the half mass radius, and is consistent with the more rigorous calculations (for both $E$ and $L$) to within a factor of $\sim5$ for mildly eccentric orbits. This number is similarly consistent with the $E$- and $L$-relaxation times of half of the cluster mass within the same factor, as shown in Figure \ref{fig:relaxation-cumulative}. Our more rigorous calculations are also based on some assumptions (e.g. isotropic velocity distribution), but importantly attempts taking into account the non-Maxwellian nature of the velocity distribution and the non-uniform spatial density. The distribution function of a Plummer sphere is proportional to $(-E)^{7/2}$ (the full expression is given in equation \ref{eq:plummer-df} in the appendix). By writing the energy at a fixed radius $r_{0}$, we can see that the distribution of velocity magnitudes at any given position is proportional to $v^2[-v^{2}-2\Phi(r_{0})]^{7/2}$ where $\Phi(r_{0})$ is a negative constant (the potential). This distribution drops to zero at the escape velocity, while the Maxwell--Boltzmann distribution has an infinite tail of high velocities. For circular orbits with $a>r_{\mathrm{h}}$, most of the conditions for the standard approximation are met and Equation (\ref{eq:chandra}) gives a very good approximation for the energy relaxation despite the somewhat different functional form of the velocity distribution. The discrepancy is most evident for eccentric orbits, and for circular orbits as well when $a<r_{\mathrm{h}}$.

Measuring the diffusion coefficients directly from $N$-body simulations is a better way to find the relaxation time as a function of the orbital elements that does not depend on any assumptions. This could in principle be done by measuring the rms change of $E$ (or $L$) denoted $\left\langle (\Delta E)^{2}\right\rangle $ of particles in a small bin in $(E,L)$-space over a short period of time $\Delta t$; the problem however is determining this $\Delta t$. The forces acting on a particle are correlated on very short timescales, and it is only on longer timescales that the random walk-like behavior is revealed. However on yet longer timescales, particles starting from a small bin in $(E,L)$-space may be scattered throughout this space, and the instantaneous rate of their energy diffusion would be affected by their new $(E,L)$ values rather than the initial ones, where we are interested in measuring the diffusion rate. This means that $\left\langle (\Delta E)^{2}\right\rangle $ as a function of $\Delta t$ is expected to be quadratic at short time intervals, transition to linear at longer intervals, and saturate to a constant value when the distribution becomes fully mixed. The problem measuring the diffusion coefficients on large values of $\Delta t$ becomes less severe for large $N$ because the longer local relaxation time everywhere means that particles deviate more slowly from their initial position in $(E,L)$-space. Furthermore, it is not guaranteed in general that there exists such an intermediate timescale which is long enough for the correlated behavior to disappear but not too long that the energies to scatter too far from the original value. This could be the case for low-$N$ systems such as open cluster, where the gravitational field is dominated by finite-number effects, and the mixing time is shorter than the orbital period. In globular clusters, however, this is unlikely to be the case as we have shown.

\citet{Theuns96} directly measured the energy diffusion coefficients, as a function of energy only, in King models from a direct-summation $N$-body simulations of up to 32k particles. He defined individual particles to be in different ``states'' between two local maxima of the $E(t)$ curve of each particle, where $\Delta E$ is defined by the difference between adjacent maxima and $\Delta t$ is the time interval between them. Doing so for both angular momentum and energy in tandem requires a larger number of particles, but easily achievable with modern computers. \citet{diemand+04} have similarly measured the mean energy relaxation times in the context of cosmological simulations, but instead of as a function of $E$, they considered different radial bins of a Hernquist model. They chose $\Delta t$ from different considerations, requiring that most particles spend most of the time interval in the same radial bin. Measurement of the diffusion time from astronomical observation is a much bigger challenge. By measuring the positions of young (bright) white dwarfs in the globular cluster 47 Tucanae, \citet{Heyl15}
were able to calculate a diffusion rate consistent with a core relaxation time of about $\sim 70\,\mathrm{Myr}$.

The results presented in Section \ref{sec:relaxation:results} can also be used to roughly estimate the diffusion coefficients (or relaxation times) for the selected orbital families, but it is only meaningful as a sanity check or as a validation of the \textsc{mocca} code. This is because unlike an $N$-body code that needs only assume Newtonian physics and gravity, the \textsc{mocca} code is essentially already programmed with scattering theory. More specifically, it is a statistical way of solving the Fokker--Planck equation, under the additional assumption of spherical spatial symmetry (velocity anisotropy however can be accommodated). Therefore, measuring the diffusion from these results would be circular. Measuring the rate of mixing from these results, however, is meaningful in the sense that it is a result of the long term stochastic behavior which is reasonably-well described by the Fokker--Planck equation. Measuring mixing from an $N$-body simulation may show additional effects not described by the Fokker--Planck equation or the approximate solution provided by \textsc{mocca}. For example, scalar resonant relaxation, if present, could be captured in the mixedness curve of $L$, but this could not be revealed in the present study, which assume spherical symmetry.

\subsection{Energy vs. angular momentum}

For processes such as loss cone refilling it is more appropriate to consider $t_{\mathrm{rx},L}$ which can differ from $t_{\mathrm{rx},E}$ and $t_{\mathrm{rh}}$ considerably under most circumstances. We found\footnote{For single-mass clusters without a central black hole.} that the difference at the half-mass radius is only a factor of a few for circular orbits, but is more than an order of magnitude for mildly eccentric orbits. However, comparing the diffusion rates of $E$ and $L$ is more difficult than comparing $t_{\mathrm{rx},E}$ and $t_{\mathrm{rx},L}$. The diffusion coefficients cannot be compared directly simply because they have different dimensionality. The diffusion coefficients are converted to relaxation times by choosing a reference $E$ and $L$, but as already discussed in Section \ref{sec:relaxation:methods}, there is some unavoidable arbitrariness to that choice. The statement we can make from the figures in Section \ref{sec:relaxation:results} is that the kinetic energy diffusion rate is faster than the angular momentum diffusion rate relative the circular angular momentum. The underlying reason can be easily understood by considering an orbit with high eccentricity (e.g. red curves in Figure \ref{fig:relaxation-elements}) and large semi-major axis. This orbit, being only weakly bound, has a small kinetic and total energy (with respect to the central potential); being almost radial, it also has small angular momentum (with respect to the circular value for that energy). Changing this orbit's average kinetic energy by its own amount (which should take approximately $t_{\mathrm{rx},E}$), would transform it generally to another weakly bound orbit, with a slightly different semi-major axis. Changing this orbit's angular momentum by the reference angular momentum, the circular angular momentum (which should take approximately $t_{\mathrm{rx},L}$), would transform it from a radial to a circular orbit. This would require many more scatterings, and thus $t_{\mathrm{rx},L}>t_{\mathrm{rx},E}$ for this kind of orbit and in general.

\subsection{Mixing}

Phase mixing occurs on the dynamical timescale for the angles, but mixing of the integrals of motion takes much longer. It has been shown \citep{Goodman+93,Hemsendorf+02} that divergence in this space is exponential on a timescale (i.e. inverse of the Liapunov exponent) proportional to the crossing time, with a factor of $(\ln\ln N)^{-1}$ leading to an extremely weak dependence on $N$. This derivation makes similar assumptions to that of scattering theory, namely that the interactions between stars are relatively discrete, separate encounters, in other words, incoherent and uncorrelated, and that the spatial distribution is uniform. Mixing in $(E,L)$-space is a different aspect of the chaotic nature of the $N$-body problem which describes the statistical spreading of constants of motion due to collisions. We quantified this process by defining \emph{mixedness}, which is a relative measure of the width of the $E$ or $L$ distribution of a subpopulation with respect to the global population. Quantitatively, it is a measure of the mean (averaging is assumed over different realizations of the initial conditions) correlation coefficient between the PDF starting from a small confined region in $(E,L)$-space and the PDF of the whole cluster. This is defined on all timescales, and its rate varies in time from very fast initially to a constant. Asymptotically at later times, mixing converges exponentially in time with a characteristic decay timescale. The mixing time scales like the relaxation time(s) with $N$ because it is driven by the same physical process of collisional diffusion.

We measured mixedness by conducting Monte Carlo simulations, which necessitated using a statistical estimator on a discrete data set, which has a bias that we attempted to correct which worked to a certain degree (i.e. on long timescales $c$ saturates at a value close to unity). Solving the Fokker-Planck equation directly, in both $E$ and $L$ with two interacting components (background and subpopulation in a certain region) may have possibly resulted in more accurate estimates of the mixedness in some aspects.

Curiously, the determination of the mixing timescale as $\sim10t_{\mathrm{rh}}$ is reminiscent of a result obtained by \citet{BarOr+12}. They derived a timescale for a small initial perturbation superimposed at a specific energy on a system at a steady state, to reach that steady state (mixing in our terminology, although they refer to this as relaxation). They showed that this timescale was equal roughly to ten times the energy diffusion time. Despite the fact that they considered a power-law cusp in the galactic center context rather than a Plummer model (which has a flat center) and used very simple Fokker--Planck analysis to derive this, it appears to be consistent with our result \citep[cf. appendix B of][]{Madigan+11}.

We demonstrated the most basic manifestation of mixing using a single-mass population in a self-gravitating Plummer sphere. The adopted definition is applicable to more general systems. In the context of spherical (i.e. globular) star clusters, multiple stellar populations are often observed. This is revealed in both spectroscopic studies which show stellar populations characterized by different chemical abundances \citep{Gratton+01,Marino+08,Carretta+09}, and photometric studies in which different populations form distinguishable sequences in color--magnitude diagrams \citep{Lee+99,Bedin+04,Piotto+07}. There is no consensus regarding the formation of such secondary population. The two leading models are ad-hoc formation of the second generation stars from the gas accumulated from the external intergalactic medium, and a minor merger of clusters with an age difference of a few hundred million years (which could be quite rare, see \citealt{Lee15}). \citealt{Hong+17} carried out numerical simulations based on these two formation scenarios and found that both of them reproduce the observed radial trend of the ratio between the stellar populations. While the spatial mixing of different populations has been studied \citep{Decressin+08,Vesperini+13,Miholics+15,HenaultBrunet+15}, the details of the isolating integrals mixing process may help to distinguish among formation models.

Finally, an important utility for the mixing time is within hybrid collisional-collisionless $N$-body codes. In this kind of scheme, the evolution of a stellar system is computed in such a way that only a fraction of the stars experiences 2-body encounters. In order to fiducially simulate such a system, requires the advance knowledge of how to divide (e.g. in $(E,L)$-space) the system into collisional and collisionless components and of how long it is possible to simulate before reassigning particles into the two groups \citep[cf. ][]{Meiron+18b}.

\acknowledgments{We thank Rainer Spurzem for stimulating discussions. We thank the referee, Douglas Heggie, for a thorough review and helpful comments, as well as Jean-Baptiste Fouvry and Jongsuk Hong for their useful feedback. This project has received funding from the European Research Council (ERC) under the European Union's Horizon 2020 research and innovation programme under grant agreement No 638435 (GalNUC) and by the Hungarian National Research, Development, and Innovation Office grant NKFIH KH-125675.}
\vfill\eject

\bibliographystyle{yahapj}
\bibliography{main}

\newpage \clearpage

\appendix

\section{Local diffusion of angular momentum magnitude}\label{sec:local-diff-L}

In this Section we write the mean square change in the angular momentum vector's magnitude during a short encounter as a function of the mean square velocity changes parallel and perpendicular to the original velocity direction. In other words, express $\langle (\Delta L)^{2}\rangle $ as a function of $\langle (\Delta v_{\parallel})^{2}\rangle $ and $\langle (\Delta v_{\bot})^{2}\rangle $. Since we are only computing the average change during a single short encounter (``local diffusion'') our expressions will depend on phase space coordinates (namely $r$, $v$ and $v_{r}$). In the next step we will integrate over them to get the orbital averaged coefficients. Also note that we are interested in the square change in the vector's magnitude, \emph{not} the square magnitude of the difference vector, thus $(\Delta L)^{2}\equiv(|\bm{L}_{2}|-|\bm{L}_{1}|)^{2}$.

We start by writing the angular momentum vector before the encounter $\bm{L}_{1}=\bm{r}\times\bm{v}_{1}$. Since the encounter occurs over a very short period, $\bm{r}$ does not change and therefore does not need to be subscripted. After the encounter, the angular momentum vector is $\bm{L}_{2}=\bm{r}\times\bm{v}_{2}$ with
\begin{equation}
\bm{v}_{2}=\bm{v}_{1}+(\Delta v_{\parallel})\hat{\bm{v}}_{1}+(\Delta v_{\bot})\hat{\bm{u}}_{1}
\end{equation}
where $\hat{\bm{v}}_{1}=\bm{v}_{1}/v_{1}$ a unit vector in the direction of $\hat{\bm{v}}_{1}$, and $\hat{\bm{u}}_{1}$ an unknown unit vector perpendicular to it. The new angular momentum following some simple algebra is
\begin{equation}
\bm{L}_{2}=\left[1+\frac{(\Delta v_{\parallel})}{v_{1}}\right]\bm{L}_{1}+(\Delta v_{\bot})(\bm{r}\times\hat{\bm{u}}_{1})
\end{equation}
and its square magnitude:
\begin{equation}
L_{2}^{2}=\left[1+\frac{(\Delta v_{\parallel})}{v_{1}}\right]^{2}L_{1}^{2}+(\Delta v_{\bot})^{2}|\bm{r}\times\hat{\bm{u}}_{1}|^{2}+2\left[1+\frac{(\Delta v_{\parallel})}{v_{1}}\right](\Delta v_{\bot})\bm{L}_{1}\cdot(\bm{r}\times\hat{\bm{u}}_{1}).
\end{equation}
The last term could be simplified as follows
\begin{equation}
\bm{L}_{1}\cdot(\bm{r}\times\hat{\bm{u}}_{1})=(\bm{r}\times\bm{v}_{1})\cdot(\bm{r}\times\hat{\bm{u}}_{1})=(\bm{r}\cdot\bm{r})(\bm{v}_{1}\cdot\hat{\bm{u}}_{1})-(\bm{r}\cdot\hat{\bm{u}}_{1})(\bm{v}_{1}\cdot\bm{r})=-r^{2}u_{1r}v_{1r}
\end{equation}
where we used that $\bm{v}_{1}\cdot\hat{\bm{u}}_{1}=0$ by definition and defined $v_{1r}$ and $u_{1r}$ as the radial components of $\bm{v}_{1}$ and $\hat{\bm{u}}_{1}$, respectively.

We are interested in the quantity
\begin{equation}
(\Delta L)^{2}=(L_{2}-L_{1})^{2}=L_{2}^{2}-2L_{2}L_{1}+L_{1}^{2}
\end{equation}
which contains the parallel and perpendicular velocity changes under a square root in the middle term. Since we only consider in those changes up to second order, we can write the $L_{2}$ as a Taylor series. The result is
\begin{align}
L_{2} & =\sqrt{\cdots}=L_{1}+\frac{L_{1}}{v_{1}}(\Delta v_{\parallel})-\frac{r^{2}v_{1r}u_{1r}}{L_{1}}(\Delta v_{\bot})-\frac{1}{2}\left[\frac{r^{4}v_{1r}^{2}u_{1r}^{2}}{L_{1}^{3}}-\frac{|\bm{r}\times\hat{\bm{u}}_{1}|^{2}}{L_{1}}\right](\Delta v_{\bot})^{2}\nonumber\\
 & +\text{higer order terms}
\end{align}
and therefore
\begin{equation}
(\Delta L)^{2} =\frac{L_{1}^{2}}{v_{1}^{2}}(\Delta v_{\parallel})^{2}+\frac{r^{4}v_{1r}^{2}u_{1r}^{2}}{L_{1}^{2}}(\Delta v_{\bot})^{2} -\frac{2r^{2}u_{1r}v_{1r}}{v_{1}}(\Delta v_{\parallel})(\Delta v_{\bot})
\end{equation}
We immediately see that the last term does not contribute to the average because $\left\langle (\Delta v_{\bot})\right\rangle =0$. Additionally, since the vector $\hat{\bm{u}}_{1}$ is independent of the change in velocity, the average is (now dropping the subscript 1)
\begin{equation}
\left\langle (\Delta L)^{2}\right\rangle =\frac{L^{2}}{v^{2}}\left\langle (\Delta v_{\parallel})^{2}\right\rangle +\frac{r^{4}v_{r}^{2}}{L^{2}}\left\langle u_{r}^{2}\right\rangle \left\langle (\Delta v_{\bot})^{2}\right\rangle .
\end{equation}
It is relatively easy to geometrically show that
\begin{equation}
u_{r}=\frac{v_{t}}{v}\cos\beta
\end{equation}
where $v_{t}^{2}=v^{2}-v_{r}^{2}$ and $\beta$ is a random angle. Since $\left\langle \cos^{2}\beta\right\rangle =1/2$ and $L=rv_{t}$ we finally get
\begin{equation}
\left\langle (\Delta L)^{2}\right\rangle =\frac{r^{2}}{v^{2}}\left[v_{t}^{2}\left\langle (\Delta v_{\parallel})^{2}\right\rangle +\frac{1}{2}v_{r}^{2}\left\langle (\Delta v_{\bot})^{2}\right\rangle \right].
\end{equation}
This result is in agreement with equation (88) of \citet{Bar-Or+16} but not in agreement with equation (21) of \citet{Spitzer+72}. The $1/2$ factor in the right term, which comes from the square cosine averaging, is not present there.

\section{Orbital averaging}\label{sec:orbital-averaging}

We finalize the calculation of the diffusion coefficients writing $\langle(\Delta v_{\parallel})^{2}\rangle$ and $\langle(\Delta v_{\bot})^{2}\rangle$ as functions of velocity with the Rosenbluth potentials as substituting into Equations (\ref{eq:local-diff-E}) and (\ref{eq:local-diff-L})
\begin{align}
\frac{\left\langle (\Delta E)^{2}\right\rangle }{\Delta t} & =\frac{8\pi\Gamma v^{3}}{3}\left[F_{4}(v)+E_{1}(v)\right]\\
\frac{\left\langle (\Delta L)^{2}\right\rangle }{\Delta t} & =\frac{8\pi\Gamma r^{2}}{3v}\left[\left(v^{2}-\frac{3}{2}v_{r}^{2}\right)F_{4}(v)+\frac{3}{2}v_{r}^{2}F_{2}(v)+v^{2}E_{1}(v)\right]
\end{align}
where $\Gamma=4\pi G^{2}m^{2}\ln\Lambda$ and
\begin{align}
F_{n}(v) & \equiv\int_{0}^{v}\left(\frac{v^{\prime}}{v}\right)^{n}f(v^{\prime})\mathrm{d}v^{\prime}\\
E_{n}(v) & \equiv\int_{v}^{\infty}\left(\frac{v^{\prime}}{v}\right)^{n}f(v^{\prime})\mathrm{d}v^{\prime}
\end{align}
are the Rosenbluth potentials. Let us use the fact that $v^{\prime}\mathrm{d}v^{\prime}=\mathrm{d}E'$ and change the variable of integration to the energy, since $f$ for the Plummer model is given as a function of only energy
\begin{align}
F_{n}(E,r) & \equiv v^{-n}(E)\int_{\Phi(r)}^{E}v^{\prime n-1}f(E^{\prime})\mathrm{d}E^{\prime}\label{eq:rosenbluth-F}\\
E_{n}(E,r) & \equiv v^{-n}\int_{E}^{\infty}v^{\prime n-1}f(E^{\prime})\mathrm{d}E^{\prime}\label{eq:rosenbluth-E}
\end{align}
given a spherically symmetric potential $\Phi(r)$ derived from $f(E)$. For a Plummer model \citep{Aarseth+74}
\begin{align}
f(E) & =\frac{24\sqrt{2}}{7\pi^{3}}\frac{r_0^2N}{G^{5}M^{5}}(-E)^{7/2}\label{eq:plummer-df}\\
\Phi(r) & =-\frac{GM}{\sqrt{r^{2}+r_0^2}}
\end{align}
where $r_0$ is the Plummer radius, $M$ is the total mass, and $G$ the gravitational constant. $f(E)=0$ outside the range $\Phi(0)<E<0$. Note that we write the distribution function such that $\int f\mathrm{d}^3\bm{x}\mathrm{d}^3\bm{v}=N$, often it is normalized to the total mass, rather than number of particles.

We can forget about the velocity dependence because both the total velocity and its radial component can be written as functions of $E,$ $L$ and $r$. The relations are
\begin{align}
v & =\sqrt{2\left[E-\Phi(r)\right]}\\
v_{r} & =\sqrt{2\left[E-\Phi(r)\right]-L^{2}/r^{2}}\label{eq:vr}
\end{align}
and similarly $v^{\prime}$ relates to $E^{\prime}$ in Equations (\ref{eq:rosenbluth-F}) and (\ref{eq:rosenbluth-E}).

Now that the local diffusion coefficients are in a form where the only phase-space co-ordinate they depend on is $r$, we can proceed to the orbital averaging for fixed $E$ and $L$. The Rosenbluth potentials (only $E_{1}$, $F_{2}$ and $F_{4}$ are needed) are tabulated for the given $E$ and radii between the pericenter $r_{\mathrm{p}}$ and the apocenter $r_{\mathrm{a}}$ (which are also functions of $E$ and $L$). The orbit-averaging integrals are \citep[chapter 2b]{Spitzer87}
\begin{align}
D_{E^{2}}(E,L) & =\frac{2}{P_r}\int_{r_{\mathrm{p}}}^{r_{\mathrm{a}}}\frac{\left\langle (\Delta E)^{2}\right\rangle }{\Delta t}\frac{\mathrm{d}r}{v_{r}}\\
D_{L^{2}}(E,L) & =\frac{2}{P_r}\int_{r_{\mathrm{p}}}^{r_{\mathrm{a}}}\frac{\left\langle (\Delta L)^{2}\right\rangle }{\Delta t}\frac{\mathrm{d}r}{v_{r}}
\end{align}
where $P_r$ is the radial orbital period (in rosette-type orbits, this is not the same as the angular period). To numerically perform the integrals, $v_{r}$ from Equation (\ref{eq:vr}) is substituted,  and we solve the potential to find $r_\mathrm{p}$, $r_\mathrm{a}$ and $P_r$ for these values of $E$ and $L$.

For circular orbits $v_{r}=0$ and there is no need to perform orbital averaging as $r$ and $v$ are constant along the orbit. The result is
\begin{align}
D_{E^{2}} & =\frac{8\pi\Gamma v^{3}}{3}\left[F_{4}(E,r)+E_{1}(E,r)\right]\\
D_{L^{2}} & =\frac{r^{2}}{v^{2}}D_{E^{2}}
\end{align}
The energy and circular velocity are related as follows to the radius
\begin{align}
E & ={\textstyle \frac{1}{2}}r\Phi^{\prime}(r)+\Phi(r)\\
v & =\sqrt{r\Phi^{\prime}(r)}
\end{align}
where $\Phi^{\prime}(r)$ is the gradient of the potential in the radial direction. The functions $E_{1}$ and $F_{4}$ still need to be evaluated as before (albeit at a single point), because they represent scattering contribution from the field particles, not orbital averaging.

\subsection{Numerical integration}

The radial velocity $v_{r}$ in the denominator in the orbital average integrals approaches zero at the apsides, causing the integrands to diverge at the integration limits. Despite the finiteness of the integrals, this poses a numerical problem which is mitigated as follows. A reasonable approximation could be made by writing the reciprocal of the problematic term as a Taylor series around each apsis and analytically calculate its integral up to a small distance $\epsilon$, where numerical integration is easier.
\begin{align}
\int_{r_{\mathrm{p}}}^{r_{\mathrm{p}}+\epsilon}v_{r}^{-1}\mathrm{d}r & =\left[\frac{2\epsilon}{-\Phi^{\prime}(r_{\mathrm{p}})+L^{2}/r_{\mathrm{p}}^{3}}\right]^{1/2}\\
\int_{r_{\mathrm{a}}-\epsilon}^{r_{\mathrm{a}}}v_{r}^{-1}\mathrm{d}r & =\left[\frac{2\epsilon}{\phantom{-}\Phi^{\prime}(r_{\mathrm{a}})-L^{2}/r_{\mathrm{a}}^{3}}\right]^{1/2}
\end{align}
We choose $\epsilon=10^{-2}(r_{\mathrm{a}}-r_{\mathrm{p}})$.

\section{Estimation of the mixedness}\label{sec:mixedness-estimation}

Here we describe the numerical procedure to estimate
\begin{equation}
c=\frac{\int p(x)p_\mathrm{bg}(x)\mathrm{d}x}{\sqrt{\int p^{2}(x)\mathrm{d}x}\sqrt{\int p_\mathrm{bg}^{2}(x)\mathrm{d}x}}
\end{equation}
from discrete data. We assume that $p_\mathrm{bg}(x)$ is a smooth function\footnote{This assumption means that the total number of particles in the system is large enough for the background distribution to be considered smooth.} and that the data $G=\{x_{1},\ldots,x_{N}\}$ is a random realization of the probability density function $p(x)$. The estimator $\hat{c}$ is calculated by the following numerical procedure. The data set $G$ is divided into $n$ bins by first throwing out the innermost and outermost $\epsilon N$ data points (this gets rid of outliers). The smallest and largest $x$-values of the remaining data points are the limits of the histogram. Then, the integral in the numerator is calculated through the trapezoidal rule where instead of $p(x)$ we use the number of data points in each bin, $p_\mathrm{bg}(x)$ is evaluated at the center of the bin. The left integral in the denominator is evaluated in the same way, and the right integral is calculated analytically or integrated numerically in some other way. This gives us a \emph{biased}
estimator $\hat{c}$. Below we describe numerical experiments we performed to attempt to correct the bias and evaluate the statistical error.

The correction factor $\zeta=c/\hat{c}$ is a function of the number of particles and the uncorrected estimator $\hat{c}$ of the projection. The number of bins used in the procedure may also play a role, but it is marginalized by choosing an optimal number of bins $n$ as a function of the number of data points $N$. The correction factor (as well as the optimal bin number) may very well depend on the exact functional form of both $p$ and $p_\mathrm{bg}$, but we assumed for the sake of the numerical experiments that both are normal distributions centered at the origin, with widths of $\sigma$ and $1$, respectively. The true value of the projection is given analytically in this case by
\begin{equation}
c=\sqrt{\frac{2\sigma}{1+\sigma^{2}}},
\end{equation}
which can be inverted to find $\sigma(c)$. The idea behind the numerical experiments is for different values of $N$ (which determines the number of bins $n$) and $c$ (which determines $\sigma$), we make many ($2^{18}$) realizations and calculated the biased estimator $\hat{c}$ for each one. This gives us a distribution of values; the correction factor $\zeta$ is the ratio $c/\left\langle \hat{c}\right\rangle $ between the real projection and the average of the biased estimators. The width of the distribution helps to determine the error.

The optimization of the number of bins to be used in the procedure is done first. It seems that the best values of $\hat{c}$ (i.e. the closest to $c$) are obtained roughly when
\begin{equation}
n=\mathrm{round}\left(1.584\ N^{0.38}\right).\label{eq:optimal-bins}
\end{equation}
The best number of bins depends on $c$ somewhat as well, but Equation (\ref{eq:optimal-bins}) marginalizes over that with some bias toward lower values of $c$. The lowest number of data points we consider is $N=32$, which gives $n=6$ bins. With $n(N)$ fixed for the procedure and the choice $\epsilon=1/64$ we proceed by generating realizations $G$ for $(c,N)$ pairs and calculate the $\hat{c}$ distribution. For each value of $N$ we choose 39 values of $c$ equally distributed between 0.05 and unity. We find that the width (representing the error) strongly depends on $N$ but weakly on $c$. At a fixed $N$, the correction factor itself as a function of $\hat{c}$ can be somewhat approximated by a 3-parameter function with the same functional form as a S\'ersic profile. We only use this fitting to evaluate the sensitivity of $\zeta$ to $N$ by looking at how the fit parameters change with $N$. We find that they vary strongly with $N$ only when the threshold to throw away another data point at the tails of the distribution is passed (e.g. between $N=191$ and $192$), therefore we make sure that those transitions are included in the grid.

The correction is generally small. The deviation of $\hat{c}$ from $c$ is the largest for the smallest number of particles ($N=32$ in our experiments) and smallest projection value ($c=0.05$ in our experiments). In this case it is still less than 15\%. It is however critical to take this projection into account for the purpose of this study, because a small bias in the mixedness value may cause a very large deviation in the mixing timescale. This ad-hoc correction is computed for the simple case where both background distribution $p_\mathrm{bg}$ and the target distribution $p$ are normal, and moreover, their centroids are the same. The results may very well depend on the functional form, but a more general or elegant derivation of the bias correction is outside the scope of this work.

\end{document}